\documentclass[
    ,final            
  ]
  {aipproc}

\layoutstyle{8x11double}

\usepackage{amsmath,amssymb,eucal,mathrsfs}

\newcommand{\be}{\begin{equation}}
\newcommand{\ee}{\end{equation}}
\newcommand{\bea}{\begin{eqnarray}}
\newcommand{\eea}{\end{eqnarray}}

\newcommand{\mc}{\mathcal}

\begin{document}

\title{A Full 24-Parameter MSSM Exploration}

\keywords      {Supersymmetry, Supersymmetric Models, Dark matter
  \quad \quad
  {\bf{Preprint}}: DAMTP-2008-75} 

\author{Shehu S. AbdusSalam}{
  address={DAMTP, Centre for Mathematical Sciences, Wilberforce Road,
  Cambridge CB3 0WA, UK} 
}

\begin{abstract}
Up until now a complete scan in all phenomenologically relevant
directions of the MSSM at the TeV scale for performing global fit has
not been done. Given the 
imminent start of operation of the LHC, this is a major gap on our
quest to discovering and understanding the physical implications of
low energy supersymmetry. The main reason for this is the large number of
parameters involved that makes it computationally extremely expensive
using the traditional methods. In this talk I demonstrate that with
advanced sampling techniques the problem is solvable. The
results from the explored 24-parameter TeV scale MSSM (phenoMSSM) are
remarkably distinct from previous studies and are independent of
models for supersymmetry breaking and mediation mechanisms. Hence they
are a more robust guide to searches for supersymmetry and dark matter.
\end{abstract}

\maketitle

\subsection{PhenoMSSM}
The $105$ free physical parameters of the MSSM with R-parity
\cite{mssm} makes a complete study of supersymmetry an impossible
task. For 
practical purposes phenomenologists had to construct models with fewer
parameters at unification scale from which RGEs were used to obtain the
lower-energy ($M_{susy}$) scale sparticle 
spectrum and properties. Most famous among this class of constructions
is mSUGRA/CMSSM  
which has just 4 parameters and a $\pm$ sign. This has been used for
providing bench-mark points for sparticle searches and
phenomenology. But the approach has important limitations. Supersymmetry 
discovery definitely requires probe of large regions, in a maximal
manner, of its parameter space. Moreover, it may be misleading if the 
models used to interpret experimental results are not realistic or not
the most general. Hence the study of MSSM in its complete 
parameters around the electroweak scale is a more natural
approach. This will be independent of supersymmetry breaking models,
hidden-sector physics, mediation mechanisms and renormalisation group
running 
interpolations.  

In the following sections I demonstrate this more natural approach by
applying advanced Monte Carlo technique, called \texttt{nested
  sampling} \cite{Skilling} which is implemented in the currently 
private code \texttt{MultiNest} \cite{Feroz:2007kg}, to fully explore
the MSSM with R-parity and minimal flavour violation (MFV) \cite{mfv}
in its entire, 24-dimensional, most phenomenologically important
parameters at the scale $M_{susy} \sim 1$ TeV -- a set-up we
call {\bf \texttt{phenoMSSM}}~\cite{phenoM}. The codes \texttt{SFitter}
and \texttt{Fittino} \cite{fitt} can reconstruct (weak-scale) MSSM
parameters from collider data but the main goal here is to explore all
the parameters, perform a global fit to current indirect data and draw
inferences for LHC and (future) LC physics. I start with definition  
of Bayes' theorem, describe how its variables were constructed for 
the {\bf phenoMSSM} and then give sample results from the exercise of
applying the theorem on the model before concluding in the last
section. 

\paragraph{{\bf Bayesian Inference in Particle Physics:}}
Bayes' theorem is at the core of the algorithm used to efficiently
explore the entire viable parameter regions of the model. It states that 
\be \label{posterior} P(\theta | D, H) = P(D | \theta, H) P(\theta|
H)/P(D | H)\ee where $P(\theta |H)$ is the parameters prior density
distribution representing the conditional probability of a set of
parameters $\theta$ given that the model or hypothesis, $H$, is
true. It states what values the model parameters are expected to
take. With some set of predictions (data or observables), $D$,
obtained from the model, $P(D |\theta, H)$ quantifies the likelihood
for the model, at the given parameter values, to be true. For a model
with $n$ parameters the $n$-dimensional integral $Z = P(D|H) = \int
P(D| \theta, H) P(\theta|H)\, d\theta$ represents the evidence for the 
model. $P(\theta | D, H)$ is the posterior probability density
function and gives a measure of how well the set of parameters
predicts the given data set, $D$. Nested sampling is
a general method for evaluating this $n$-dimensional integral by
converting it to a 1-dimensional integral over a unit interval
\cite{Skilling}. The sampling procedure also produces the posterior
probability distribution eqn.~(\ref{posterior}) as by-product. 

\paragraph{{\bf{PhenoMSSM parameters, $\theta$:}}}
The soft supersymmetry breaking part of the MSSM Lagrangian density   
have contributions from different types of interactions, ${\mc L}_{\rm
  soft} = {\mc L}_{\rm gauginos} + {\mc L}_{\rm sfermions} + {\mc
  L}_{\rm trilinear} + {\mc L}_{\rm higgs}$, and sources the 105
supersymmetric free parameters. In order to suppress CP-violation and
FCNC real soft terms, diagonal sfermions masses and trilinear
couplings, and 1st/2nd generation squark masses and slepton masses
degeneracies were assumed. $A_t$, $A_b$ and $A_\tau$ are the most
important trilinear couplings but we also include $A_e = A_\mu$
because it is relevant for $(g-2)_\mu$ computation
\cite{Martin:2001st}. All the other trilinear couplings and neutrino
masses are set to zero. This way the total number of free parameters
becomes $20$. Adding to these the $4$ most important SM
``nuisance'' parameters: $\{m_t, m_b(m_b)^{\overline{MS}},
\alpha_{em}(m_Z)^{\overline{MS}}, \alpha_{s}(m_Z)^{\overline{MS}}\}$,
makes a total of $24$ physical parameters in
{\bf{\texttt{phenoMSSM}}}. These are listed and described in 
Table~\ref{tab.param}. 
\begin{table}[!ht] 
\begin{tabular}{cl}
\hline
\tablehead{1}{r}{b}{Parameter} & \tablehead{1}{r}{b}{Description} \\
\hline
$M_1$, $M_2$, $M_3$       & Bino, Wino and Gluino masses      \\
$m_{\tilde e_L} = m_{\tilde \mu_L}$    & 1st/2nd generation $L_L$
slepton masses \\  
$m_{\tilde \tau_L}$  & 3rd generation $L_L$ slepton mass\\   
$m_{\tilde e_R} = m_{\tilde \mu_R}$    & 1st/2nd generation $E_R$
    sleptons masses\\  
$m_{\tilde \tau_R}$  & 3rd generation $E_R$ slepton mass\\   
$m_{\tilde u_L} = m_{\tilde d_L} = $\\ $m_{\tilde c_L} = m_{\tilde s_L}$ &
    1st/2nd generation $Q_L$ squark masses \\   
$m_{\tilde t_L} = m_{\tilde b_L}$ & 3rd generation $Q_L$ squark masses\\
$m_{\tilde u_R} = m_{\tilde c_R}$ & 1st/2nd generation $U_R$ squark masses \\  
$m_{\tilde t_R}$ & 3rd generation $U_R$ squark mass  \\  
$m_{\tilde d_R} = m_{\tilde s_R}$ & 1st/2nd generation $D_R$ squark masses\\  
$m_{\tilde b_R}$ & 3rd generation $D_R$ squark mass \\  
$A_{t,b,\tau}$ & top, b- and $\tau$- quark trilinear couplings   \\
$A_e = A_\mu$ & $\mu$ and $e$ trilinear couplings \\
$m_{H_{1,2}}$ & up- and down-type Higgs doublet masses \\
$\tan \beta$ & scalar doublets {\bf vev}s ratio \\ 
$m_t$ & top quark pole mass  \\
$m_b(m_b)^{\overline{MS}}$ & b-quark mass \\
1/$\alpha_{em}(m_Z)^{\overline{MS}}$ & electromagnetic coupling
    constant \\ 
$\alpha_{s}(m_Z)^{\overline{MS}}$ & strong coupling constant \\ 
\hline
\end{tabular}
\caption{The 24 parameters of {\b{\texttt{phenoMSSM}}}
  set-up.}\label{tab.param}
\end{table}
Representing them in a 24-dimensional vector $\theta$,
the combined prior for the model is \be \pi(\theta) = P(\theta | H) =
\pi(\theta_1) \, \pi(\theta_2) \, \ldots \pi(\theta_{24}). \ee

\paragraph{{\bf{Set of Observables, $D$:}}}
We use high precision electroweak and B-physics collider observables
and the dark matter relic density from WMAP5 results (all shown in
Table~\ref{tab.observ}) to study the viable parameter regions of the
{\bf phenoMSSM}. Its predictions for these observables were
obtained from the 24 input parameters via 
\texttt{SOFTSUSY2.0.17}~\cite{Allanach:2001kg} for producing the MSSM
spectrum; 
\texttt{micrOMEGAs2.1}~\cite{micromegas} for computing neutralino dark
matter relic density, the branching ratio $BR(B_s \rightarrow \mu^+
\mu^-)$ and the anomalous magnetic moment of the muon $(g-2)_\mu$;
\texttt{SuperIso2.0}~\cite{Mahmoudi:2007vz} for predicting the Isospin
asymmetry in the decays $B \rightarrow K^* \gamma$ and $BR(b \rightarrow s
\gamma)$ with all NLO supersymmetric QCD and NNLO SM QCD contributions
included; and \texttt{susyPOPE}~\cite{spope} for computing $W$-boson
mass $m_W$, the effective leptonic mixing angle variable $\sin^2
\theta^{lep}_{eff}$, and the total $Z$-boson decay width, $\Gamma_Z$, at
two loops in the dominant MSSM parameters. 
\begin{table}[!ht] 
\begin{tabular}{cll}
\hline
\tablehead{1}{r}{b}{Observable}
 & \tablehead{1}{r}{b}{Mean value} & \tablehead{1}{r}{b}{Uncertainty}  \\
\hline
$m_W$ & 80.398 GeV  & 0.0025 GeV\\
$\Gamma_Z$ & 2.4952 GeV & 0.0023 GeV\\
$\sin^2\, \theta_{eff}^{lep}$ & 0.23149 & 0.000173\\
$\delta a_\mu \times 10^{10}$ & 29.5 & 8.8 \\ 
$Br(b \rightarrow s \gamma) \times 10^{4}$ & 3.55 & 0.72 \\ 
$m_h$ & 114.4 GeV & lower limit \\
$Br(B \rightarrow \mu^+ \mu^-)$ & $5.8 \times 10^{-8}$ & upper limit
\\ 
$R_{\Delta M_{B_s}}$ & 0.85  & 0.11 \\ 
$R_{Br(B_u \rightarrow \tau \nu)}$ & 1.2589 & 0.4758\\
$\Delta_{0-}$ & 0.0375 & 0.0289 \\
$\Omega_{CDM} h^2$ & 0.1143 & 0.02\\
\hline
\end{tabular}
\caption{A summary of the 11 observables. $\Omega_{CDM} h^2$ error is
  inflated to 0.02 to accommodate theoretical
  uncertainties.}\label{tab.observ}    
\end{table}
These physical observables derived from the model parameters form the
data set, $D$. For each element, $D_i$, in $D$ the likelihood $L(\theta) =
P(D_i|\theta, H)$ was calculated. Assuming that the observables are
independent then the combined likelihood for the model is  
\be L(\theta) = \prod_{i} \, \left(2\pi \sigma_i^2\right)^{-1/2}
\exp\left[- (O_i - \mu_i)^2/2\sigma_i^2 \right] \ee  where $O_i$ is
the predicted value of the $i$th observable with $i = 1, 2, 3, \ldots,
24$ and $\sigma_i$ its corresponding standard error.  

\begin{figure}
  \includegraphics[height=.15\textheight]{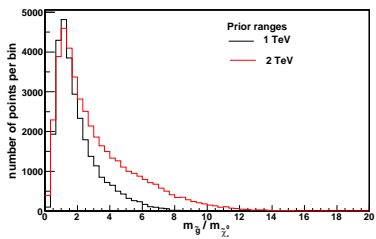} 
  \caption{ppd of gluino to neutralino mass ratio showing most
  probable phenoMSSM values around 2 and severely disfavouring
  mSUGRA/CMSSM and AMSB.} \label{fig.mratio}
\end{figure}

With all the above Bayesian inference parameters set and ready the
\texttt{Multinest} code was employed for (guided) sampling of the $24$
parameters. At each parameter-space point the values were passed in
SUSY Le Houches Accord (SLHA) format~\cite{slha} to the different
particle physics software used for predicting the physical
observables. The predictions were then checked against experimental
values with (non)deviations quantified by the likelihood
function. Next, the likelihoods modulate the parameter prior
probabilities to produce the Bayesian evidence and posterior
probability distributions for the model. Two prior probability
density ranges, 1 TeV and 2 TeV,  were used for the purely
supersymmetric parameters (the first 20 listed in
Table~\ref{tab.param}) with the gaugino masses and trilinear couplings
allowed to take both positive and negative values. The slepton and
squark masses were bounded from below at $100$ GeV. The SM
parameters were taken as Gaussian noise around their mean
experimental values: $m_t = 172.6 \pm 1.4$, $m_b(m_b)^{\overline{MS}}
= 4.2 \pm 0.07$, $1/\alpha_{em}(m_Z)^{\overline{MS}} = 127.918 \pm
0.018$ and $\alpha_{s}(m_Z)^{\overline{MS}} = 0.1172 \pm 0.002$. 
\paragraph{{\bf{Results}}}
\begin{figure}[ht]
  \begin{tabular}{ll}
   \begin{minipage}[t]{3.5cm}
    \includegraphics[width=1.2\textwidth, height=.168\textheight]{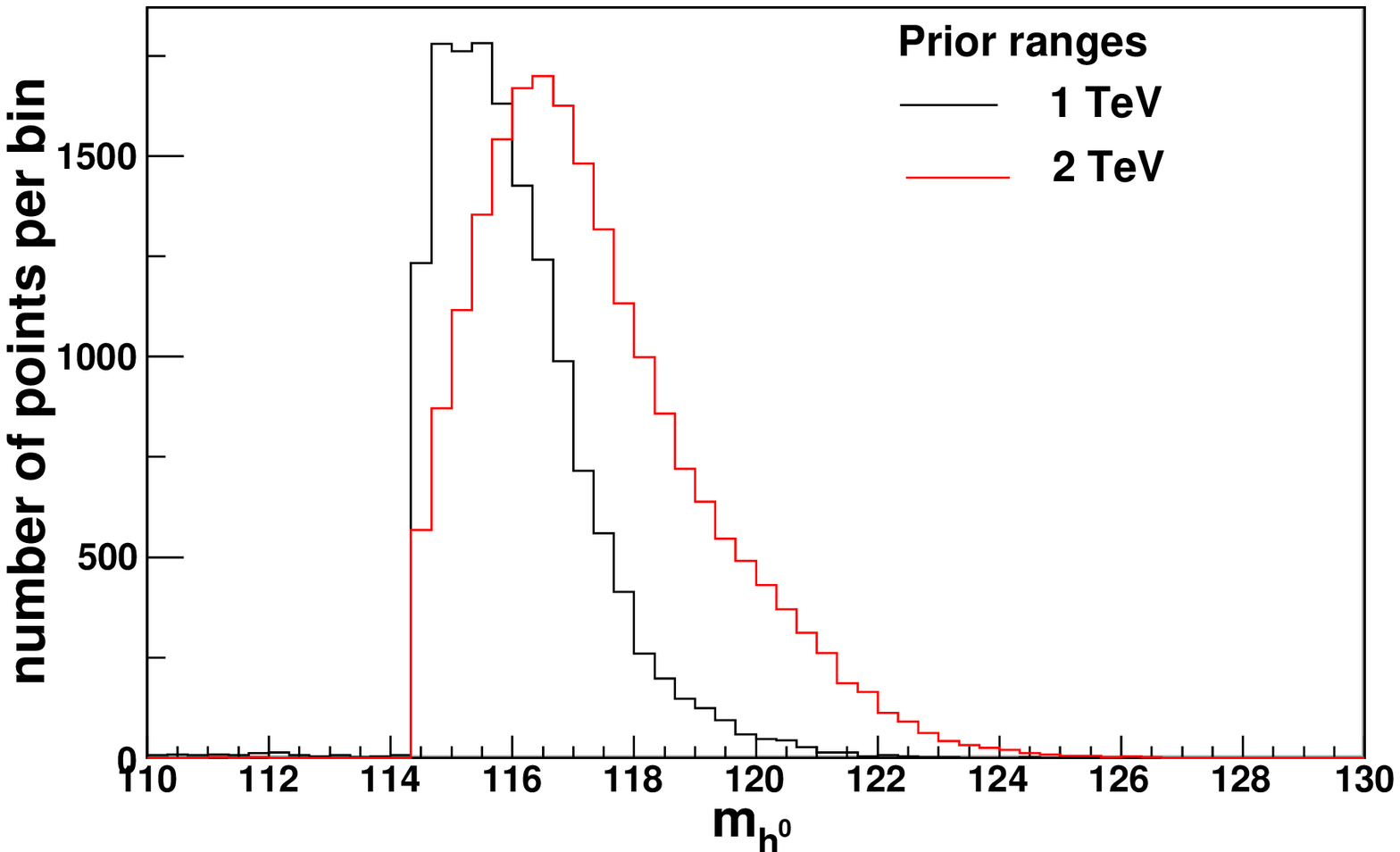}
   \end{minipage}
&
  \begin{minipage}[t]{3.5cm}
    \includegraphics[width=1.2\textwidth,
    height=.16\textheight]{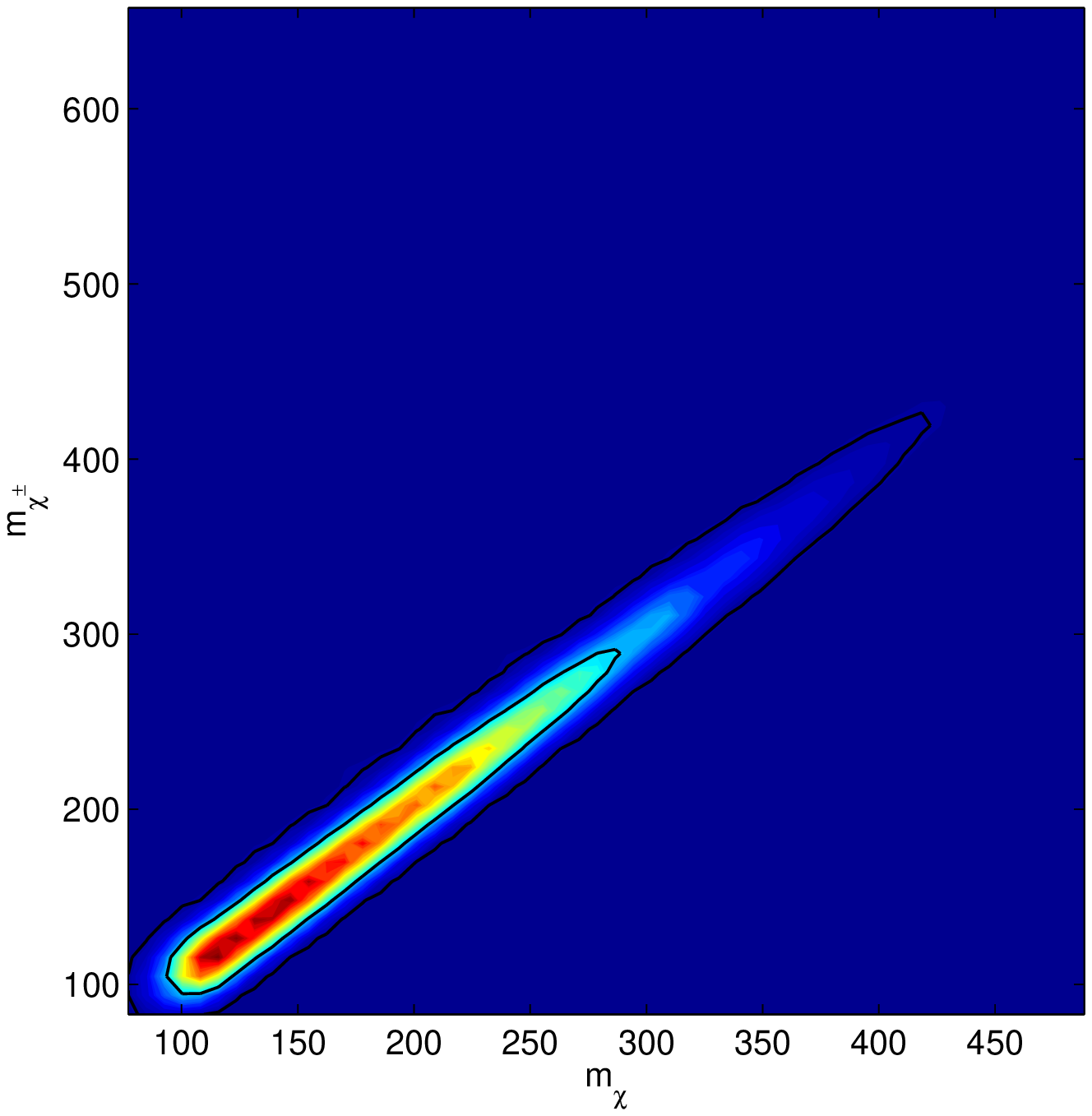}
   \end{minipage}
   \end{tabular}
  \caption{{\bf Left}: Posterior probability distribution (ppd) for
  $m_{h^0}$. {\bf Right}: Marginalised ppd
  showing chargino co-annihilation; colour coded: 0 for blue to
  red/black for maximum.} \label{fig.higgs}
\end{figure}
Here I give some results from the exploration exercise, more results
and analysis on the global fit to data is in progress~\cite{phenoM}. 
The results are quite robust under change of parameter prior
ranges. Figure~\ref{fig.higgs} shows that $m_{h^0}$ is most likely
around $115$ to $117$ GeV, just above the LEP limit. The
different neutralino dark matter (DM) annihilation mechanisms at early
times of the universe that leads to the value of its relic density
today are important in defining different phenomenological regions in
model parameter space. A 
profound feature of the {\bf phenoMSSM} is that it shows lots
of chargino co-annihilation as shown in Figure~\ref{fig.higgs}. The
gluino to neutralino mass ratio is an 
interesting quantity that characterises different models of
supersymmetry (breaking.) For instance mSUGRA(AMSB) with
predominantly bino(wino) LSP has $m_{\tilde{g}}/m_{\tilde{\chi}^0_1}
\approx6(9)$~\cite{nilles}. The mirage mediation~\cite{mirage} and
the LARGE volume~\cite{LV} scenarios have the characteristic ratio less
than 6 and between 3 to 4 respectively. Figure~\ref{fig.mratio} shows
the ppd of this  quantity, providing discrimination between the {\bf
  phenoMSSM} and the other models. This feature is robust to prior
parameter ranges. 
\paragraph{{\bf{Conclusion}:}}
With advanced Bayesian technique it is possible to fully explore
weak-scale MSSM in all of its phenomenologically relevant parameter
directions. Doing this is very important since low energy
supersymmetry is the main focus of the LHC supersymmetry-search
experiments. Hence such low energy approach as demonstrated here will
be a better guide and much more realistic. The {\bf phenoMSSM} has a
characteristic $m_{\tilde g}/m_{\tilde{\chi}^0_1} 
\lesssim 2$ and shows the existence of chargino-neutralino (DM)
co-annihilation. The whole procedure can be applied to other BSM
constructions such as those with additional CP-violating and 
FCNC sources and non-zero neutrino masses. Moreover, using the
evidence value, $Z$, the technique is a powerful tool for
model comparisons~\cite{Feroz:2008wr}.

\paragraph{{\bf Acknowledgements}}
Thanks to B.C.~Allanach and F.~Quevedo for comments on the notes. The
computations were done on COSMOS, UK's national cosmology
supercomputer at DAMTP, Cambridge. The author is supported by the
Gates Cambridge Trust scholarship.

\end{document}